\documentclass{nature_fig}

\usepackage[english]{babel}
\usepackage{bbold}
\usepackage{bbm}
\usepackage[pdftex]{graphicx}
\usepackage{latexsym,amsmath,verbatim}
\usepackage{color}
\usepackage{rotating}
\usepackage{multirow}
\usepackage{color}

\usepackage{amssymb,amsfonts,amsmath}
\usepackage{booktabs}
 
\usepackage[normalem]{ulem}

\renewcommand{\(}{\left(}
\renewcommand{\)}{\right)}

\newcommand{\tr}[1]{\text{Tr}\(#1\)}
\renewcommand{\(}{\left(}
\renewcommand{\)}{\right)}

\newcommand{\heading}[1]{{\vspace{0.25truecm}\noindent\textbf{#1.}}}

\usepackage{xcolor}
\definecolor{RoyalBlue}{HTML}{4169e1}
\definecolor{ForestGreen}{HTML}{228b22}

\usepackage[colorlinks = true,
            linkcolor = blue,
            urlcolor  = blue,
            citecolor = blue,
            anchorcolor = blue]{hyperref}
 
\usepackage{xcolor}
\definecolor{mypink2}{RGB}{219, 48, 122}

\usepackage{subfigure}
\usepackage{comment}

\usepackage{todonotes}

\title{Multiscale statistical physics of the Human--SARS-CoV-2 interactome}

\author{Arsham Ghavasieh$^{1}$, Sebastiano Bontorin$^{1,2}$, Oriol Artime$^{1}$, Manlio De Domenico$^{1\ast}$}

\begin{document}

\maketitle

\begin{affiliations}
\item Fondazione Bruno Kessler, Via Sommarive 18, 38123 Povo, Italy
\item Department of Physics, University of Trento, Via Sommarive 14, 38123 Povo (TN), Italy
\end{affiliations}
\noindent Corresponding author: mdedomenico@fbk.eu
\date{}

\baselineskip24pt


\begin{abstract}
Protein-protein interaction (PPI) networks have been used to investigate the influence of SARS-CoV-2 viral proteins on the function of human cells, laying out a deeper understanding of COVID--19 and providing ground for drug repurposing strategies. However, our knowledge of (dis)similarities  between this one and other viral agents is still very limited. Here we compare the novel coronavirus PPI network against 45 known viruses, from the perspective of statistical physics. 
Our results show that classic analysis such as percolation is not sensitive to the distinguishing features of viruses, whereas the analysis of biochemical spreading patterns allows us to meaningfully categorize the viruses and quantitatively compare their impact on human proteins. Remarkably, when Gibbsian-like density matrices are used to represent each system's state, the corresponding macroscopic statistical properties measured by the spectral entropy reveals the existence of clusters of viruses at multiple scales. Overall, our results indicate that SARS-CoV-2 exhibits similarities to viruses like SARS-CoV and Influenza A at small scales, while at larger scales it exhibits more similarities to viruses such as HIV1 and HTLV1.
\end{abstract}

\pagebreak

The COVID--19 pandemic, with global impact on multiple crucial aspects of human life, is still a public health threat in most areas of the world. Despite the ongoing investigations aiming to find a viable cure, our knowledge of the nature of disease is still limited, especially regarding the similarities and differences it has with other viral infections. On the one hand, SARS-CoV-2 shows high genetic similarity to SARS-CoV~\cite{Andersen2020} --- the virus causing 2003 coronavirus outbreak --- and its infection shares a number of symptoms with some other respiratory diseases, such as flu caused by Influenza virus. On the other hand, according to a number of studies~\cite{costanzo2010genetic,Young2020,Cao2020,Choy2020}, drugs usually used to treat different infection types, like AIDS caused by Human Immunodeficiency Virus (HIV), can effectively mitigate COVID--19, suggesting an unexplored parallel between the function of other viruses and SARS-CoV-2. Characterizing these (dis)similarities can result in a deeper understanding of the novel coronavirus and facilitate the search for reliable treatments. 

With the rise of network medicine~\cite{barabasi2011network,ivanov2016focus,zhou2014human,silverman2012network,goh2007human,halu2019multiplex}, methods developed for complex networks analysis have been widely adopted to efficiently investigate the interdependence among genes, proteins, biological processes, diseases and drugs~\cite{Sonawane2019}. Similarly, they have been used for characterizing the interactions between viral and human proteins in case of SARS-CoV-2 ~\cite{Gordon2020, cui2020structural, vandelli2020structural}, providing insights into the structure and function of the virus~\cite{Estrada2020} and identifying drug repurposing strategies~\cite{gysi2020network,ray2020predicting}. However, a comprehensive comparison of SARS-CoV-2 against other viruses, from the perspective of network science, is still missing.

Here, we use statistical physics to analyze 45 viruses, including SARS-CoV-2. We consider the virus-human protein-protein interactions (PPI) as an interdependent system with two parts, human PPI network targeted by viral proteins. In fact, due to the large size of human PPI network, its structural properties barely change after being merged with viral components. Consequently, we show that percolation analysis of such interdependent systems provides no information about the distinguishing features of viruses. Instead, we model the propagation of perturbations from viral nodes through the whole system, using bio-chemical and regulatory dynamics, to obtain the spreading patterns and compare the average impact of viruses on human proteins. Finally, we exploit Gibbsian-like density matrices, recently introduced to map network states, to quantify the impact of viruses on the macroscopic functions of human PPI network, such as von Neumann entropy. The inverse temperature $\beta$ is used as a resolution parameter to perform a multiscale analysis. We use the above information to cluster together viruses and our findings indicate that SARS-CoV-2 groups with a number of pathogens associated with respiratory infections, including SARS-CoV, Influenza A and Human Adenovirus (HAdV) at the smallest scales, more influenced by local topological features. Interestingly, at larger scales, it exhibits more similarity with viruses from distant families such as HIV1 and Human T- cell Leukemia Virus type 1 (HTLV1).

Our results shed light on the unexplored aspects of SARS-CoV-2, from the perspective of statistical physics of complex networks, and the presented framework opens the doors for further theoretical developments aiming to characterize structure and dynamics of virus-host interactions, as well as  grounds for further experimental investigation and potentially novel clinical treatments.


\section*{Results}
\begin{figure}
    \centering
    \includegraphics[width=0.65\textwidth]{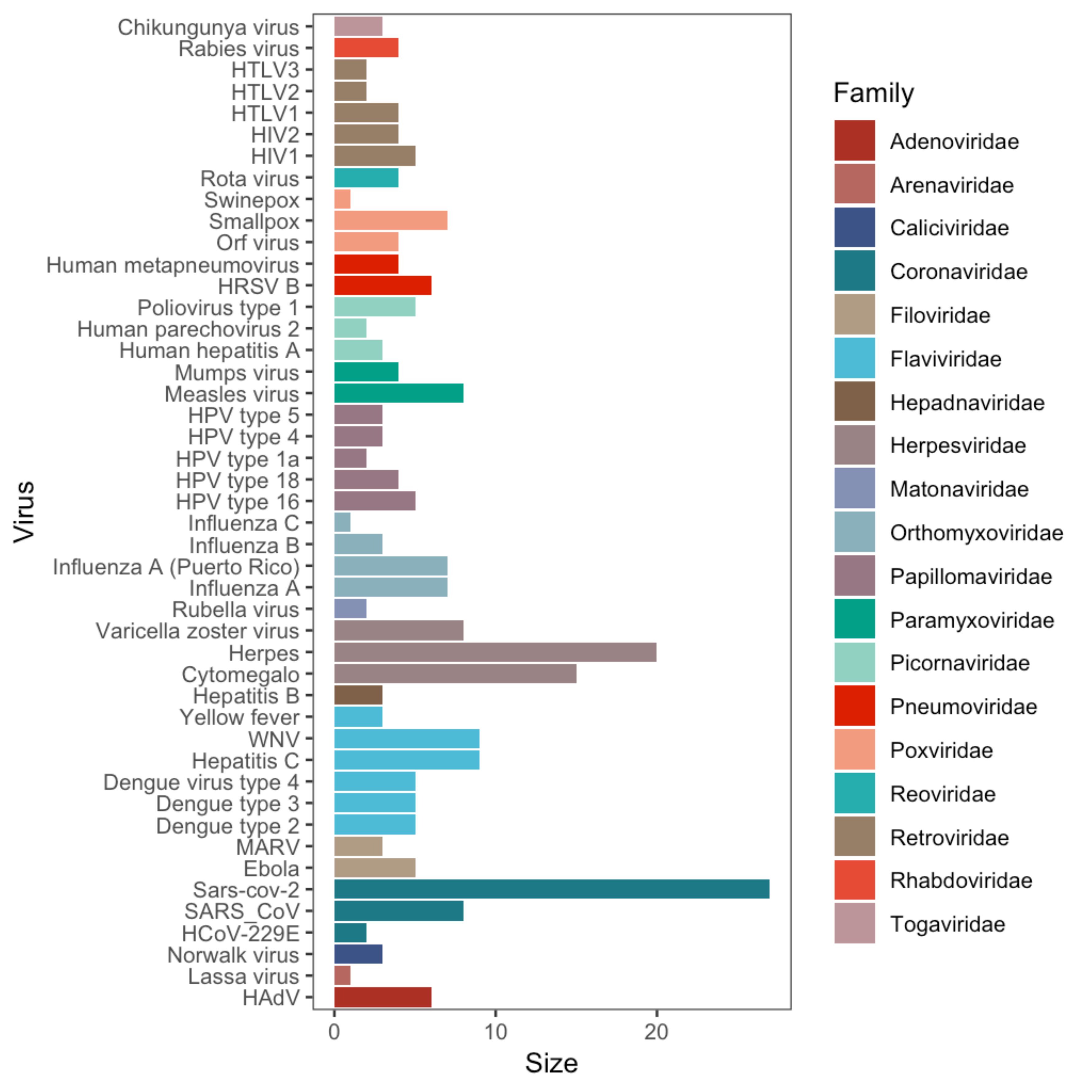}
    \caption{\label{fig:virus_families}\textbf{Virus information summary.} The 45 viruses used in this study are shown against their size, in terms of viral proteins, and coloured by their official family classification.}
\end{figure}

\begin{figure}
    \centering
    \includegraphics[width=\textwidth]{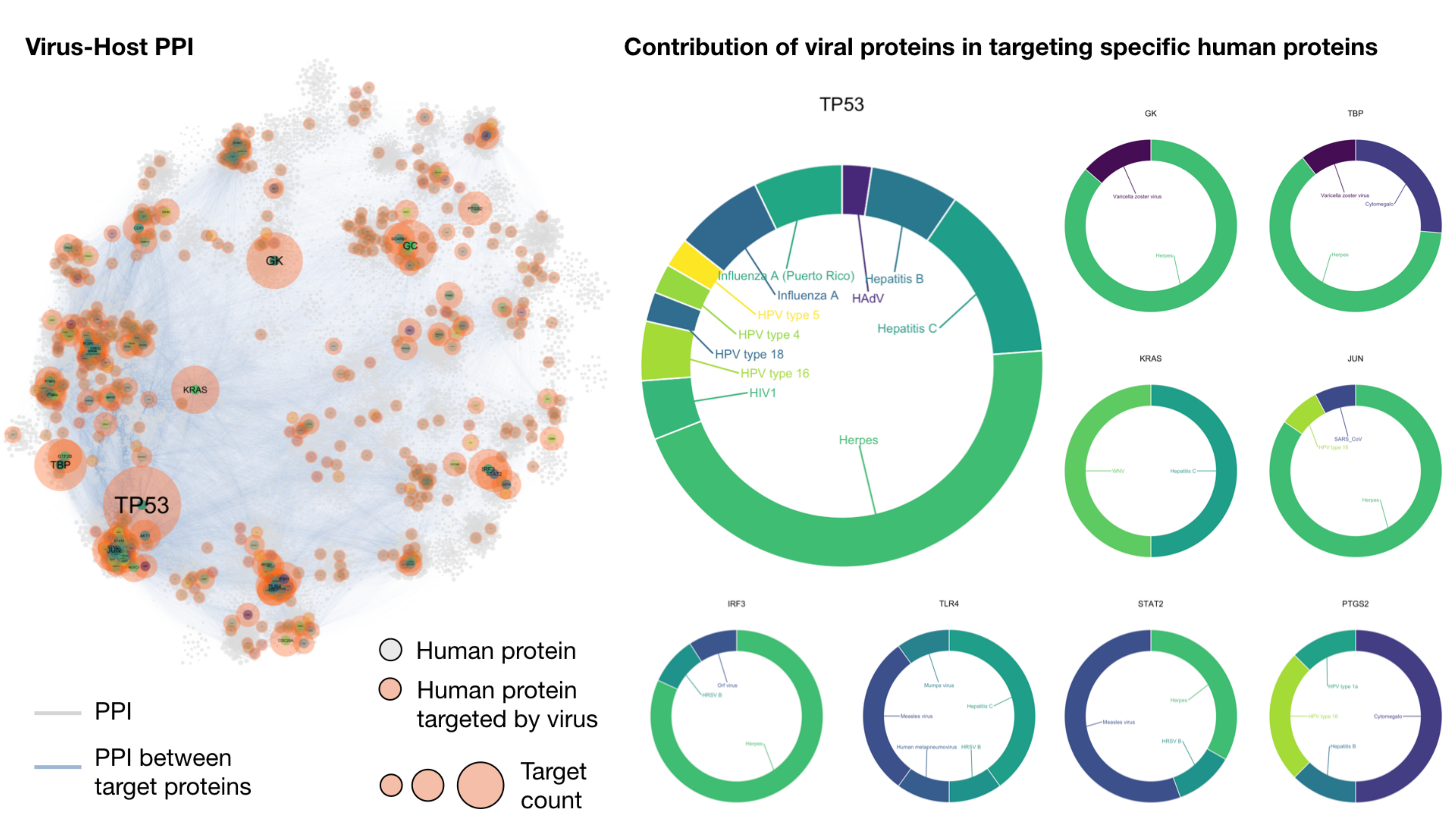}
    \caption{\label{fig:virus_contribution}\textbf{Virus-host interactome as an interdependent network.} BIOSTR Human PPI used in this study, this obtained from data fusion of two comprehensive public repositories, namely STRING and BIOGRID (see the text for details). The network consists of $N=19,945$ proteins linked by $|E|=737,668$ edges, and the largest connected component (99.8\% nodes, 99.6\% edges) is shown. Proteins targeted by viruses are highlighted in two ways. On the one hand, markers of distinct size identify targeted proteins: bigger the marker larger the number of times a protein is targeted by viruses in our data set. On the other hand, distinct colored markers of constant size encode distinct viruses (45 in total, including SARS-CoV-2): on the right-hand side the same color scheme is used to show the contribution of each virus to the most frequently targeted proteins.}
\end{figure}

Here, we use data regarding the viral proteins and their interactions with human proteins for 45 viruses (see Methods and Fig.~\ref{fig:virus_families}). To obtain the virus-human interactomes, we link the data to the BIOSTR Human PPI network ($19,945$ nodes and $737,668$ edges) built from data fusion of two comprehensive public repositories (see Methods and Fig.~\ref{fig:virus_contribution}).

\textbf{Percolation of the interactomes.} Arguably, the simplest conceptual framework to assess how and why a networked system loses its functionality is via the process of percolation~\cite{callaway2000network}. Here, the structure of interconnected systems is modeled by a network $\mathcal{G}$ with $N$ nodes, which can be fully represented by an adjacency matrix $\mathbf{A}$ ($A_{ij}=1$ if nodes $i$ and $j$ are connected, it is 0 otherwise).
\begin{figure}[!ht]
    \centering
    \includegraphics[width=0.5\textwidth]{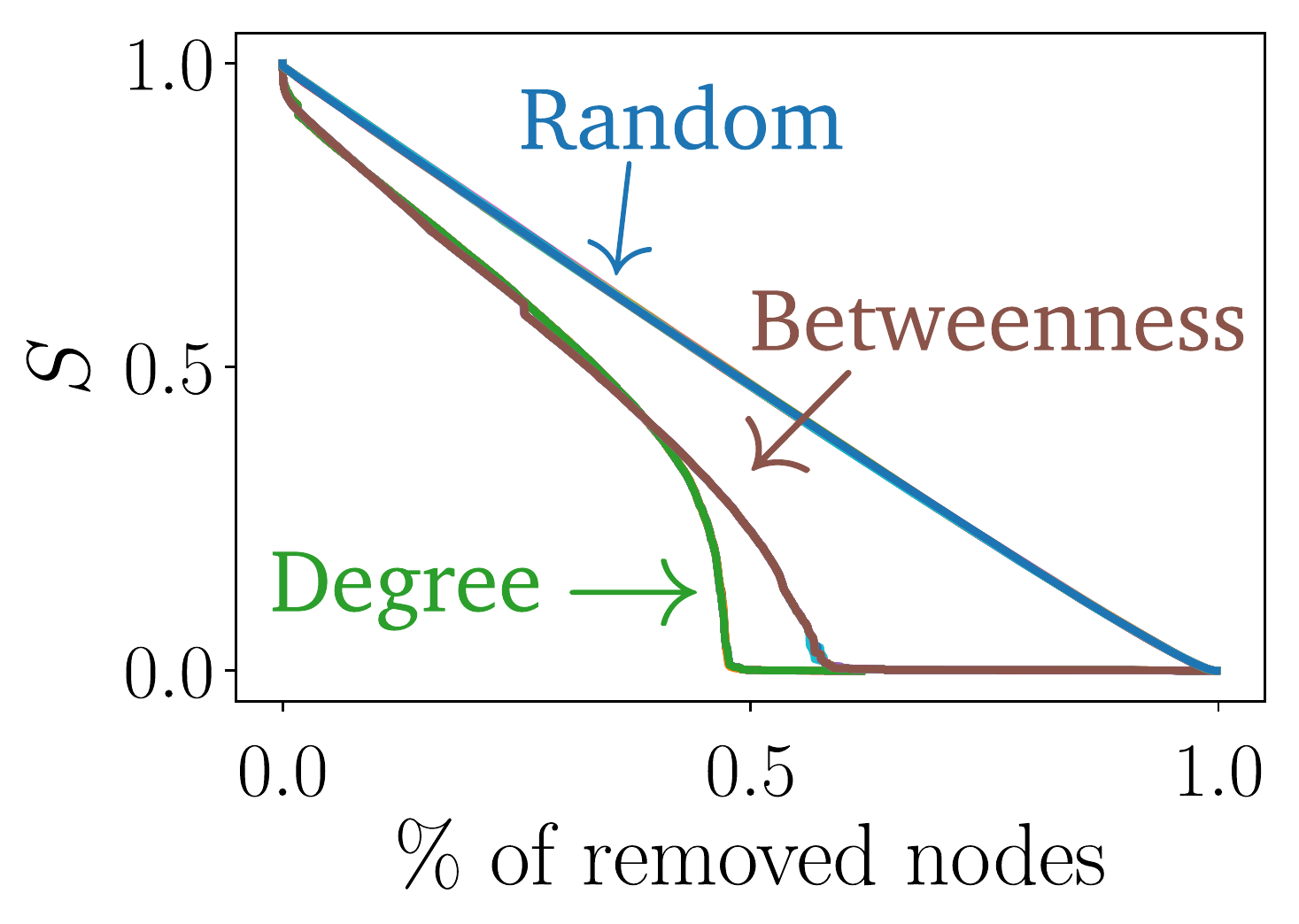}
    \caption{\textbf{Percolation analysis}. Normalized size of the largest connected component for three distinct network interventions (see the text for details). All $45$ virus are considered for each strategy and shown accordingly: note that they overlap into a common curve characteristic for each attack strategy. The degrees are recalculated after each removal, while the node ranking for betweenness is computed at the beginning of the intervention and kept fixed during the removals. \label{fig:FigPerc}}
\end{figure}
In the classical version of percolation analysis, one removes a randomly chosen fraction of nodes or links (depending on the application domain) from the original network and computes statistical and geometrical properties of the remaining subnetworks, such as the size of the largest connected component, the size distribution of isolated clusters or correlation functions, among others~\cite{stauffer2018introduction}. This point of view assumes that, as a first approximation, there is an intrinsic relation between connectivity and functionality: when the node removal occurs, the more capable of remaining assembled a system is, the better it will perform its tasks. Hence, we have a quantitative way to assess the robustness of the system. If one wants to single out the role played by a certain property of the system, instead of selecting the nodes randomly, they can be sequentially removed following that criteria. For instance, if we want to find out what is the relevance of the most connected elements on the functionality, we can remove a fraction of the nodes with largest degree~\cite{albert2000error, cohen2001breakdown}. Technically, the criteria can be whatever metric that allows us to rank nodes, although in practical terms topologically-oriented protocols are the most frequently used due to their accessibility, such as degree, betweenness, etc. Therefore percolation is, at all effects, a topological analysis, since its input and output are based on structural information. 

In the past, the usage of percolation has been proved useful to shed light on several aspects of protein-related networks, such as in the identification of functional clusters~\cite{zhang2006identification} and protein complexes~\cite{wang2010identifying}, the verification of the quality of functional annotations~\cite{gilks2005percolation} or the critical properties as a function of mutation and duplication rates~\cite{kim2002infinite}, to name but a few. Following this research line, we perform the percolation analysis to all the PPI networks to understand if this technique brings any information that allows us to differentiate among viruses. The considered protocols are the random selection of nodes, the targeting of nodes by degree -- i.e., the number of connections they have -- and their removal by betweenness centrality -- i.e., a measure of the likelihood of a node to be in the information flow exchanged through the system by means of shortest paths. We apply these attack strategies and compute the resulting (normalized) size of the largest connected component $S$ in the network, which serves as a proxy to the remaining functional part, as commented above. This way, when $S$ is close to unity the function of the network has been scarcely impacted by the intervention, while when $S$ is close to $0$ the network can no longer be operative. The results are shown in Fig.~\ref{fig:FigPerc}. Surprisingly, for each attacking protocol, we observe that the curves of the size of the largest connected component neatly collapse in a common curve. In other words, percolation analysis completely fails at finding virus-specific discriminators. Viruses do respond differently depending on the ranking used, but this is somehow expected due to the correlation between the metrics employed and the position of the nodes in the network. 

We can shed some light on the similar virus-wise response to percolation by looking at topological structure of the interactomes. Despite being viruses of diverse nature and  causing such different symptomatology, their overall structure shows a high level of similarity when it comes to the protein-protein interaction. Indeed, for every pair of viruses we find the fraction of nodes $f_N$ and fraction of links $f_L$ that simultaneously participate in both. Averaging over all pairs, we obtain that $ f_N = 0.9996 \pm 0.0002$ and $ f_L = 0.9998 \pm 0.0007$. That means that the interactomes are structurally very similar, so the dismantling ranks. If purely topological analysis is not able to differentiate between viruses, then we need more convoluted, non-standard techniques to tackle this problem. In the next sections we will employ these alternative approaches.

\textbf{Analysis of perturbation propagation.} PPI networks represent the large scale set of interacting proteins. In the context of regulatory networks, edges encode dependencies for activation/inhibition with transcription factors. PPI edges
can also represent the propensity for pairwise binding and the formation of complexes. The analytical treatment of these processes is described via Bio-Chemical dynamics ~\cite{eberhard2000, Maslov2007} and Regulatory dynamics ~\cite{alon2006introduction}.
In Bio-Chemical (Bio-Chem) dynamics, these interactions are proportional to the product of concentrations of reactants, thus resulting in a second-order interaction, forming dimers. Protein concentration $X_{i}$ ($i=1,2,...,N$) is also dependent on its degradation rate $B_{i}$ and the amount of protein synthesized at a rate $F_{i}$.
The resulting Law of Mass Action: $\dot x_{i} = F_{i} - B x_{i} + \sum\limits_{j=1j}^{N} A_{ij} x_{i}x_{j}$ summarizes the formation of complexes and degradation/synthesis processes that occur in a PPI.
Regulatory dynamics can be instead characterized by an interaction with neighbors described by a Hill function that saturates at unity: $\dot x_{i} = - x_{i} +  \sum\limits_{j=1}^{N} A_{ij} x_{j}^{h}/(1 + x_{j}^{h})$, the Michaelis-Menten (M-M) model.

\begin{figure}[ht!]
    \centering
    \includegraphics[width=0.5\textwidth]{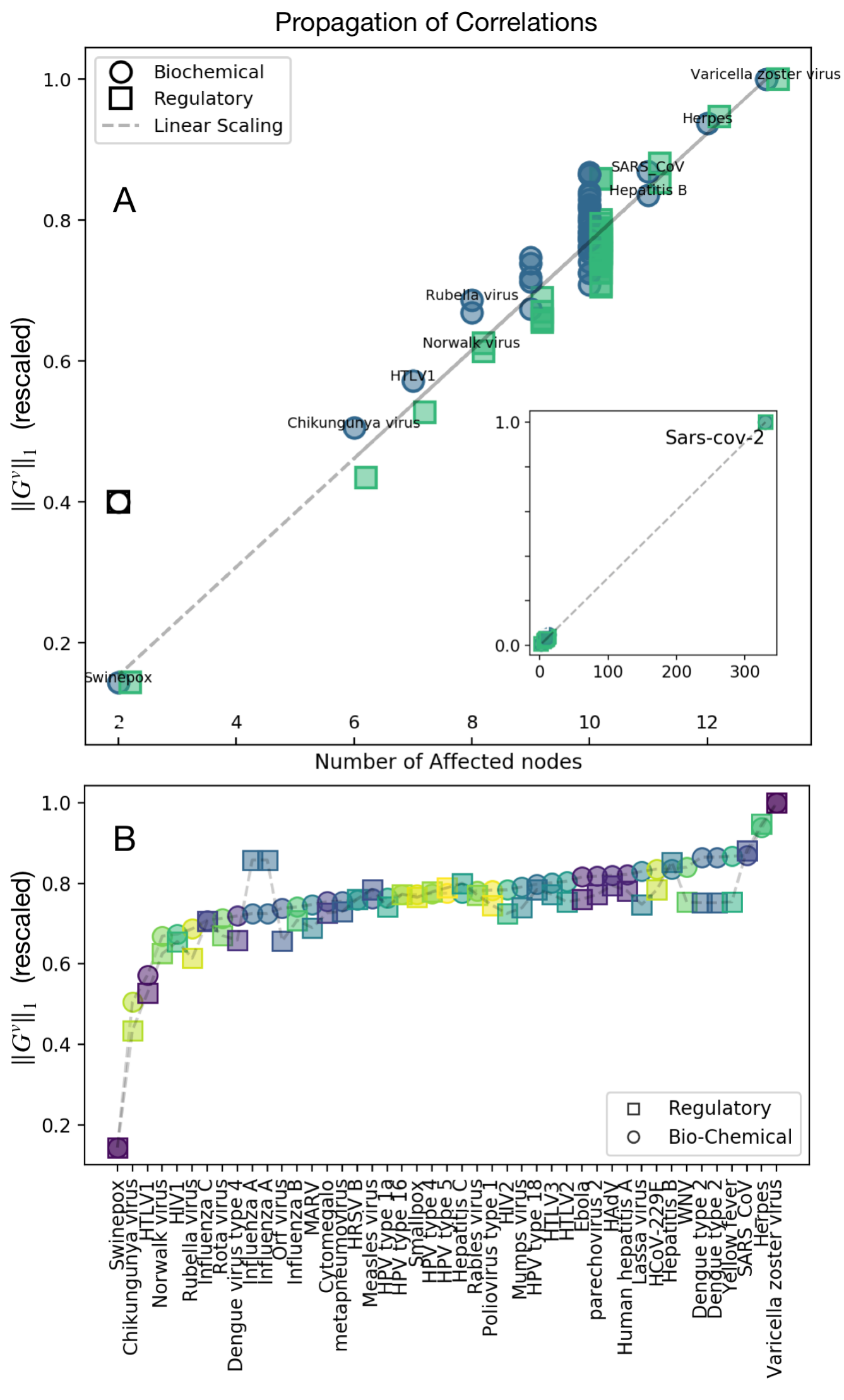}
    \caption{\textbf{Cumulative Correlation} \small{\textbf{A}: Linear Scaling of the $\vec{G}^{v}$ norm; we find that the amount of correlation distributed is linearly proportional to the number of sources of perturbation, as suggested by scaling laws. Note that in the case of SARS-CoV-2 (inset) the number of sources is 332, the largest one observed in our data set.
    \textbf{B}: Rescaled  $\| \vec{G}^{v} \|_{1}$ vs Viruses. Here SARS-CoV-2 is not shown for clarity.
    In both panels Bio-Chem and M-M norms are rescaled for better comparison by division with their relative maximum value = $\displaystyle\max_{v \in \text{viruses}} \| \vec{G}^{v} \| $}}
    \label{corr_figure}
\end{figure}

\emph{Propagation of a perturbation from virus interactions.} Viral proteins interact with specific nodes (human proteins) of the reconstructed interactome, which we refer to as a set of affected nodes $\mathcal{V}$. In the context of Regulatory and Biochemical dynamics we model the effect of viral interaction as an inhibition in the activity of the affected nodes or decrease in the concentration. The effect is simulated by introducing a negative constant perturbation at the steady state concentration/activity $x_{i} \rightarrow x_{i} - \alpha x_{i}$,  $\forall i \in \mathcal{V}$ (e.g., $\alpha = 0.2$) and tracking its propagation to the rest of the network by solving the corresponding set of the coupled equations. For a M-M--like model (with $h=1$) it leads to

\begin{equation}
    \begin{cases}
\dot x_{i} = 0 & \text{if } i \in \mathcal{V}\\
\dot x_{i} = - x_{i} +  \displaystyle\sum_{j=1}^{N} A_{ij} \frac{x_{j}}{1 + x_{j}} & \text{otherwise} \\
 \end{cases}
 \label{MM}
\end{equation}

In the context of the study of signal propagation, recent works have introduced the definition of network Global Correlation Function ~\cite{Barzel2009,Barzel2013} as
\begin{equation}
    G_{ij} = \left| \frac{ d x_{i}/ x_{i}}{ d x_{j}/x_{j}} \right|.
\end{equation}
Ultimately, the idea is that constant perturbation brings the system to a new steady state $x_{i} \rightarrow x_{i} + dx_{i}$, and $ d x_{i}/ x_{i}$ quantifies the magnitude of the response of node $i$ from the perturbation in $j$. This allows also the definition of measures such as Impact ~\cite{Barzel2013} of a node as $I_{i} = \sum_{j} A_{ij} G_{ij}^{T}$ describing the response of $i$'s neighbors to its perturbation. 
Interestingly, it was found that these measures can be described with power laws of degrees ($I_{i} \approx k_{i}^{\phi}$), via universal exponents dependent on the dynamics underlying ODEs allowing to effectively describe the interplay between topology and dynamics. In our case, $\phi =0$ for both processes, therefore the perturbation from $i$ has the same impact on neighbors, regardless of its degree. We exploit the definition of $G_{ij}$ to define the vector $\vec{G}^{v}$ of perturbations of concentrations induced by the interaction with the virus $v$, where the $k$--th entry is given by~\cite{Barzel2013}
\begin{equation}
    G_{k}^{v} = \frac{1}{\alpha}\left| \frac{dx_{k}}{x_{k}} \right |.
\end{equation}

The steps we follow to asses the impact of the viral nodes in the human interactome via the microscopic dynamics are described next. We first obtain the equilibrium states of human interactome by numerical integration of equations. Then, for each virus, we compute the system response from perturbations starting in $\forall i \in \mathcal{V}$ which is eventually encoded in $\vec{G}^{v}$. Finally, we repeat these steps for both the Bio-Chem and M-M models. 
The amount of correlation generated is a measure of the impact of the virus on the interactome equilibrium state. We estimate it as the Euclidean 1-norm of the correlation vectors $\| \vec{G}^{v} \|_{1}  = \sum_{i} |G_{i}^{v}|$, which we refer to as \textit{Cumulative Correlation}. The results are presented in Fig.~\ref{corr_figure}. 

\begin{figure}
    \centering
    \includegraphics[width=0.75\textwidth]{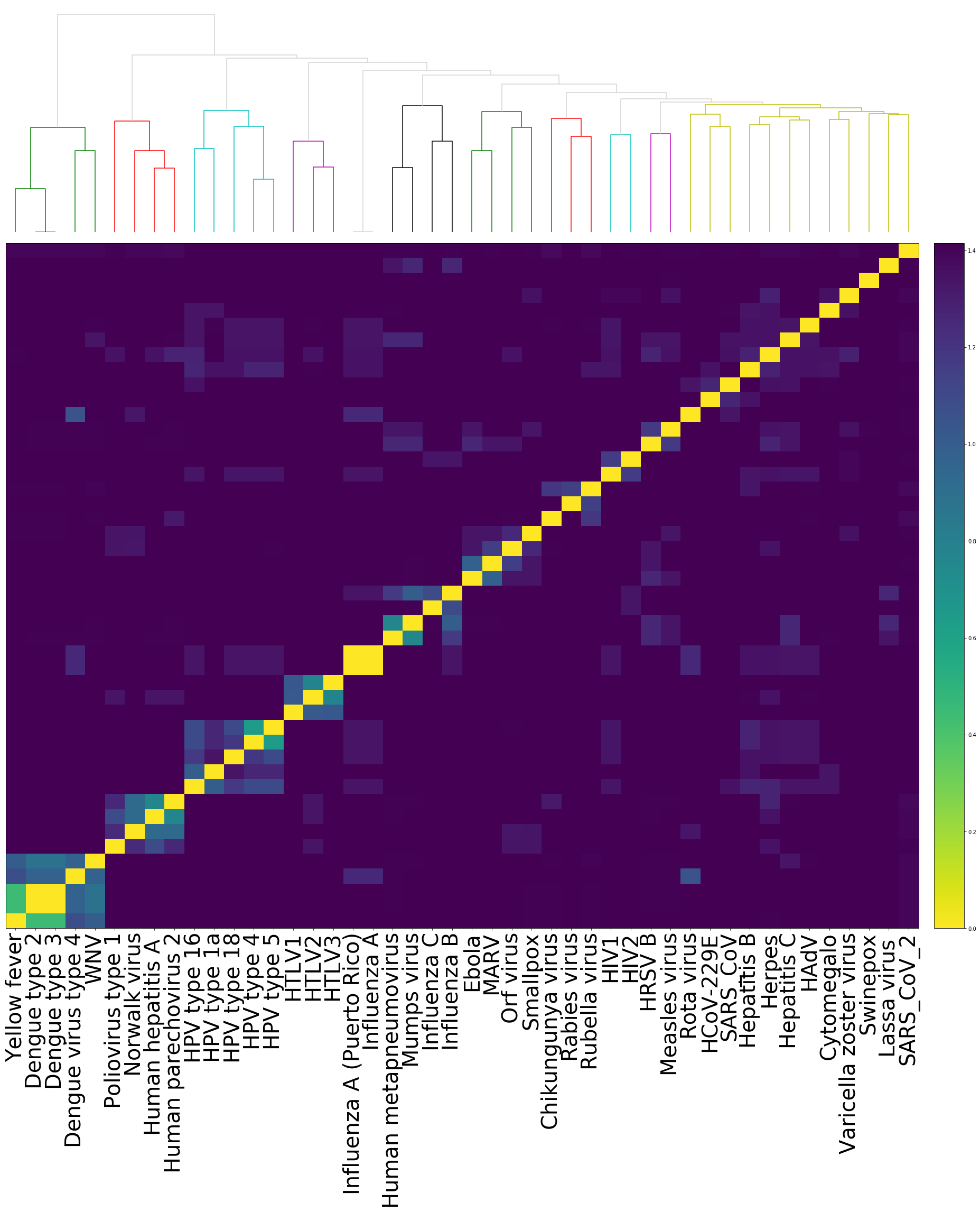}
    \caption{\textbf{Similarity analysis.} Vectors $\vec{G}_{v}$ describe the pattern of perturbations generated on the full network, and can therefore be used as a \textit{perturbation state} generated by virus $v$. We seek similarities in perturbation states by evaluating pairwise Euclidean distances between normalized $\vec{G}_{v}$.  Flaviviridae, Papillomaviridae, Picornaviridae and HTLV viruses in the Retroviridae families are well clustered together, as shown in the dendrogram. Other clusters are less sharp and depend on the choice of threshold, here chosen arbitrarily as 60 \% of the maximum distance between the clusters. \label{fig:patterns_clus}}
\end{figure}

By allowing for multiple sources of perturbation, the biggest responses in magnitude will come from direct neighbors of these sources, making them the dominant contributors to $\|\vec{G}^{v}\|_{1}$. With $I_{i}$ not being dependent on the source degree, these results support the idea that with these specific forms of dynamical processes on the top of the interactome, the overall impact of a perturbation generated by a virus is proportional to the amount of human proteins it interacts with.

Results shown in Fig.~\ref{fig:patterns_clus} highlight that propagation patterns strongly depend on the sources (i.e., the affected nodes $\mathcal{V}$), and strong similarities will generally be found within the same family and for viruses that share common impacted proteins in the interactome. Conversely, families and viruses with small (or null) overlap in the sources exhibit low similarity and are not sharply distinguishable. To cope with this, we adopt a rather macroscopic view of the interactomes in the next section.

\textbf{Analysis of spectral information.} We have shown that the structural properties of human PPI network does not significantly change after being targeted by viruses. Percolation analysis seems ineffective in distinguishing the specific characteristics of virus-host interactomes while, in contrast, the propagation of biochemical signals from viral components into human PPI network has been shown successful in assessing the viruses in terms of their average impact on human proteins. Remarkably, the propagation patterns can be used to hierarchically cluster the viruses, although some of them are highly dependent on the choice of threshold (Fig.~\ref{fig:patterns_clus}). In this section, we use statistical physics of complex networks to go beyond the microscopic details provided by propagation patterns and analyze the macroscopic features of virus-human PPI networks.

\begin{figure}
    \centering
    \includegraphics[width=\textwidth]{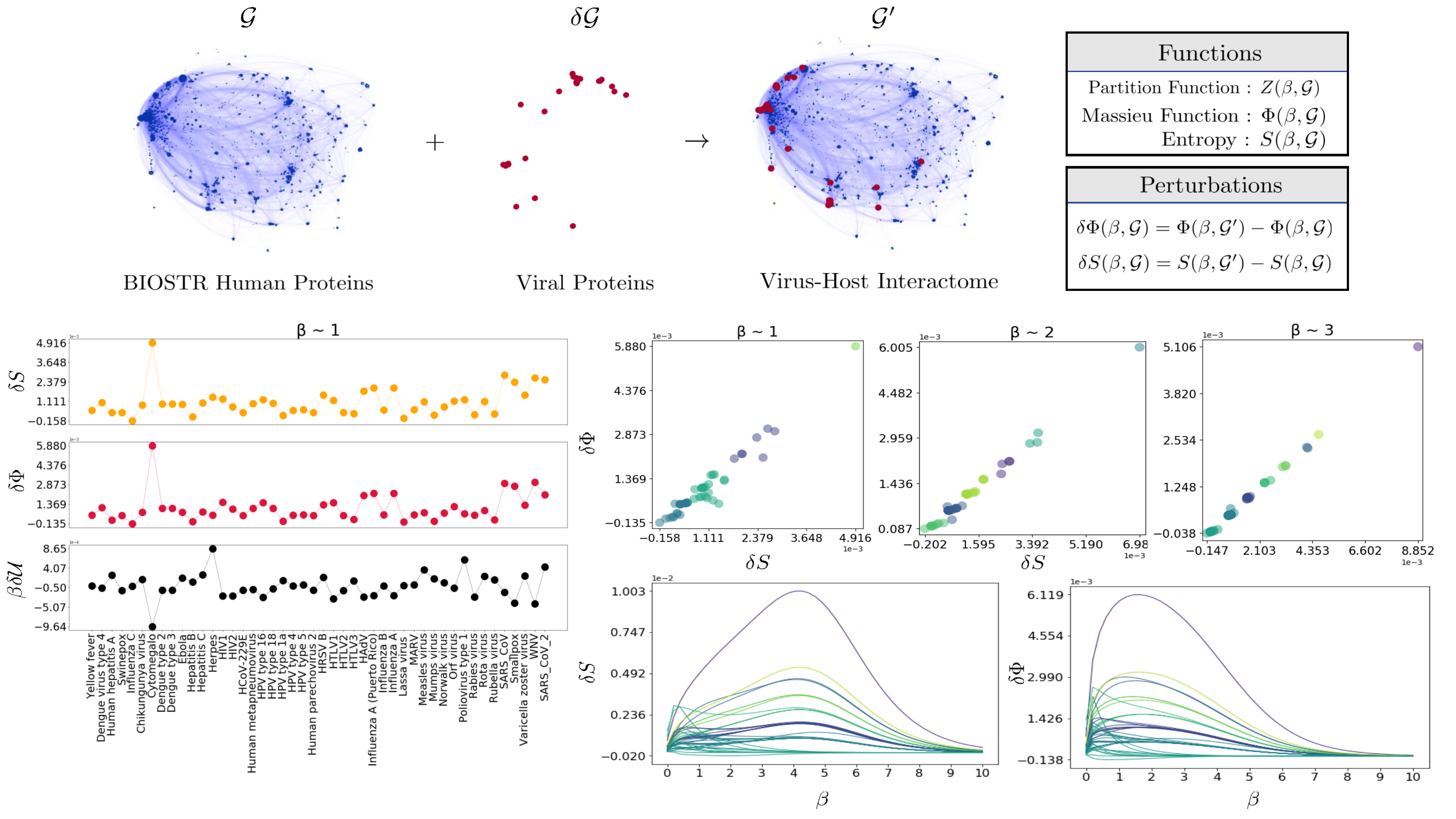}
    \caption{\label{fig:HVinterdependent}\textbf{Perturbation analysis of the virus-host interactome.} On the top, the BIOSTR human interactome  $\mathcal{G}$ is targeted by viral proteins, considered as microscopic perturbations  ($\delta\mathcal{G}$), to build the virus-human interactome $\mathcal{G}'$. Here, the SARS-CoV-2 interactome is shown, while excluding the 10\% human proteins with the highest degree, for clarity. The interdependence is reflected in the macroscopic functions of the network (right-hand side tables), including the thermodynamic and dynamic features considered in this study. Interacting with the viral nodes perturbs the macroscopic properties of the human PPI network (bottom-left panel), captured by the von Neumann entropy $\mathcal{S}(\beta,\mathcal{G})$, the Massieu function $\Phi(\beta,\mathcal{G})$ and the energy $\beta \mathcal{U}(\beta,\mathcal{G})$. Based on the von Neumann entropy and Massieu function perturbations, k-means algorithm is used to cluster the viruses at different scales corresponding to distinct choices of $\beta$ (bottom-right panel). The trajectories at the bottom indicate how the perturbation changes with $\beta$ and their colors are set by the clustering plot given at $\beta\approx3$. }
\end{figure}

A variety of methods have been introduced to analyze the information content of complex networks~\cite{Cimini2019,Radicchi_2020}. Since networks can be viewed as collections of entangled entities, a density matrix can be used to describe their state as in quantum statistical mechanics. While some choices of the density matrix have been shown to be unphysical~~\cite{Passerini2008,de2015structural}, Gibbsian-like density matrices have been successfully used to define spectral entropy~\cite{spectral2016,Biamonte2019} and estimate the information content of empirical complex networks at multiple scales, with applications ranging from transportation systems~\cite{ATransport2020} to the human microbiome~\cite{spectral2016} and the human brain~\cite{Nicolini2020}. 

The density matrix is defined in terms of the combinatorial Laplacian matrix $\mathbf{L}=\mathbf{D}-\mathbf{A}$, where $\mathbf{D}$ is defined as $D_{ii}=k_{i}\delta_{ij}$ and $k_{i}=\sum\limits_{j}A_{ij}$ denotes the degree of $i$--th node. The Laplacian matrix governs the diffusion dynamics on top of the network and is involved in the linear stability analysis of many complex dynamics, such as synchronization. Here we use the Gibbs state given by
\begin{eqnarray}
\mathbf{\rho}(\beta,\mathcal{G}) = \frac{e^{-\beta \mathbf{L}}}{\tr{e^{-\beta \mathbf{L}}}},
\end{eqnarray}
which is defined in terms of the propagator of a diffusion process on top of the network, normalized by the partition function $Z(\beta,\mathcal{G}) = \tr{e^{-\beta \mathbf{L}}}$, which has an elegant physical meaning in terms of dynamical trapping for diffusive flows~\cite{ATransport2020}. Consequently, the counterpart of Massieu function --- also known as free entropy --- in statistical physics can be defined for networks as
\begin{equation}
    \Phi(\beta,\mathcal{G})= \log{Z(\beta,\mathcal{G})}.
\end{equation}
Note that a low value of the Massieu function indicates high information flow between the nodes. The von Neumann entropy can be directly derived from the Massieu function by
\begin{equation}
    \mathcal{S}(\beta,\mathcal{G})= - \beta \partial_{\beta} \Phi(\beta,\mathcal{G}) + \Phi(\beta,\mathcal{G}),
\end{equation}
encoding the information content of graph $\mathcal{G}$. Finally, the difference between von Neumann entropy and the Massieu function follows
\begin{equation}
\label{eq:energy}
    \mathcal{S}(\beta,\mathcal{G})-\Phi(\beta,\mathcal{G})=\beta \mathcal{U}(\beta,\mathcal{G}),
\end{equation}
where $\mathcal{U}(\beta,\mathcal{G})$ is the counterpart of internal energy in statistical physics. In the following, we use the above quantities to compare the interactomes corresponding to different virus-host interactomes. In fact, as the number of viral nodes is much smaller than the number of human proteins, we model each virus-human interdependent system $\mathcal{G}'$ as a perturbation of the large human PPI network $\mathcal{G}$ (See Fig.~\ref{fig:HVinterdependent}). 

After considering the viral perturbations, the von Neumann entropy, Massieu function and the energy of the human PPI network change slightly. The magnitude of such perturbations can be calculated as explained in Fig.~\ref{fig:HVinterdependent}, for von Neumann entropy and Massieu function, while the perturbation in internal energy follows their difference $\beta \delta \mathcal{U}(\beta,\mathcal{G}) = \delta\mathcal{S}(\beta,\mathcal{G})-\delta\Phi(\beta,\mathcal{G})$, according to Eq.~\ref{eq:energy}. The parameter $\beta$ encodes the propagation time in diffusion dynamics, or equivalently an inverse temperature from a thermodynamic perspective, and is used as a resolution parameter tuned to characterize macroscopic perturbations due to node-node interactions at different scales, from short to long range~\cite{Ghavasieh2020UnravelingTE}. 

Based on the perturbation values and using k-means algorithm, a widely adopted clustering technique, we group the viruses together (see Fig.~\ref{fig:HVinterdependent}, Tab.~\ref{tab:beta1} and Tab.~\ref{tab:beta3}). At small scales, SARS-CoV-2 appears in a cluster with a number of other viruses causing respiratory illness, including SARS-CoV, Influenza A and HAdV. However, at larger scales, it exhibits more similarity with HIV1, HTLV1 and HPV type 16.

\begin{table}
\centering
\begin{tabular}{|c|c|}
\toprule
   \textbf{Cluster ID} & \textbf{Viruses} \\\hline
            1 &\begin{tabular}[c]{@{}c@{}} 
            SARS-CoV-2 - SARS-CoV - Influenza A - Influenza A(Puerto Rico) \\  HAdV  - Smallpox - WNV    
            \end{tabular}         \\ \hline
            2 &\begin{tabular}[c]{@{}c@{}} 
            HPV type 1a - HPV type 4 - HPV type 5 -
            Hepatitis A - Hepatitis B \\ Influenza B - Influenza C - HTLV2 - HTLV3 \\
            Mumps virus - Rubella virus - HCoV-229E  - Human parechovirus 2 \\  Swinepox - Lassa virus - Yellow fever -  MARV - Rabies virus\\
            \end{tabular} \\ \hline
            3 &\begin{tabular}[c]{@{}c@{}} Dengue type 2- Dengue type 3 - Dengue type 4 - Hepatitis C\\ 
            HIV1 - HIV2 - HTLV1 - HPV type 16 - HPV type 18\\
            Human metapneumovirus - HRSV B \\
            Chikungunya virus -  Ebola -  Herpes \\ Measles virus - Norwalk virus - Orf virus \\ Poliovirus type 1 - Rota virus -  Varicella zoster virus \end{tabular} \\ \hline
            4 &\begin{tabular}[c]{@{}c@{}} Cytomegalo \end{tabular} \\ \hline
\end{tabular}
\caption{\label{tab:beta1}The summary of clustering results at small scales ($\beta \approx 1$ from Fig.\ref{fig:HVinterdependent}) is presented. Remarkably, at this scale, SARS-CoV-2 groups with a number of respiratory diseases including SARS-CoV, Influenza A and HAdV.}
\end{table}

\begin{table}
\centering
\begin{tabular}{|c|c|}
\toprule
   \textbf{Cluster ID} & \textbf{Viruses} \\ \hline
            1 &\begin{tabular}[c]{@{}c@{}} 
            SARS-CoV-2 - HIV1 - HTLV1 - HPV type 16 \\ 
             \end{tabular} \\  \hline
            2 & \begin{tabular}[c]{@{}c@{}}  Influenza A - Influenza A (Puerto Rico) - HAdV \end{tabular} \\ \hline
            3 &\begin{tabular}[c]{@{}c@{}} 
            Dengue type 2 - Dengue type 3 - Dengue type 4 - HIV2 \\ Human metapneumovirus - HRSV B - Orf virus - Varicella zoster virus \\ HPV type 18 \end{tabular} \\ \hline
            4 & \begin{tabular}[c]{@{}c@{}} 
            Human hepatitis A - Influenza C - Hepatitis B - Herpes \\ 
            HPV type 1a - HTLV3 - Lassa virus - Mumps virus \\ 
            Poliovirus type 1 - Rubella virus \end{tabular} \\ \hline
            5 &\begin{tabular}[c]{@{}c@{}} 
            SARS-CoV - Smallpox  \end{tabular} \\ \hline
            6 & \begin{tabular}[c]{@{}c@{}} Yellow fever - Swinepox - Chikungunya virus - Ebola \\ Hepatitis C - HCoV-229E - HPV type 4 - HPV type 5 \\ Human parechovirus 2 - HTLV2 - Influenza B - MARV \\ Measles virus - Norwalk virus - Rabies virus - Rota virus \end{tabular} \\\hline
            7 & \begin{tabular}[c]{@{}c@{}} 
            WNV \end{tabular} \\ \hline
            8 &\begin{tabular}[c]{@{}c@{}} Cytomegalo \end{tabular} \\ \hline
\end{tabular}
\label{tab:my_label}
\caption{\label{tab:beta3}The summary of clustering results at large scales ($\beta \approx 3$ from Fig.\ref{fig:HVinterdependent}) is presented. Here, SARS-CoV-2 shows higher similarity to HIV1, HTLV1 and HPV type 16.}
\end{table}

\section*{Discussion}

Comparing COVID--19 against other viral infections is still a challenge. In fact, various approaches can be adopted to characterize and categorize the complex nature of viruses and their impact on human cells. 

In this study, we used an approach based on statistical physics to analyze virus-human protein-protein interactions outlining 45 different viral infections. Our findings suggest that viral components have negligible effect on the structural properties of the human PPI network, therefore classical topological analysis is not able to unravel distinctive patterns. Alternatively, we analyzed the propagation of perturbations from viral components into the human PPI network, allowing us to compare viruses in terms of their average impact on human proteins and their corresponding biochemical spreading patterns. While this analysis provides microscopic details about the complex interactions between viral and human proteins, it is not sufficient to identify distinctive patterns relating SARS-CoV-2 and existing viruses. Finally, from the analysis of macroscopic features in terms of information-theoretic and thermodynamic-like quantities, such as the von Neumann entropy, the Massieu function and internal energy, we have been able to cluster the viruses across multiple scales, determined by the resolution parameter $\beta$. According to our results, SARS-CoV-2 shows similarity with SARS-CoV, influenza A and a number of other viruses causing respiratory infections, as one could plausibly expect. At larger scales, where the interplay between the topology of virus-host interaction and information flow dynamics becomes more relevant, SARS-CoV-2 appears to be more similar to viruses like HIV1 and HTLV1.
As mentioned earlier, the response of COVID--19 patients has been shown to be positive to drugs used for treating HIV infections, like lopinavir-ritonavir~\cite{costanzo2010genetic,Young2020,Cao2020,Choy2020}. Our findings, in parallel with such a clinical evidence, suggests a rather unexplored relationship between SARS-CoV-2 and HIV1, motivating further theoretical and experimental investigations.

Overall, our framework opens the doors for further analyses of viral agents from the perspective of statistical physics, highlighting the sensitivity of macroscopic functions, such as spectral entropy, to small variations across interaction networks and, more specifically, virus-host interactomes. Even though the analysis of perturbation propagation patterns lacks the same sensitivity, according to our results it provides microscopic details about the interactions between viral and human proteins that complement the macroscopic view, together enhancing our understanding of this novel coronavirus from a new perspective which can provide a mathematical ground for the exploration of further clinical treatments and biological understanding.

\section*{Methods}

\textbf{Overview of the data set.} The human interactome used in this study combines protein-protein interactions (PPI) from two of the largest repository publicly available to date, namely STRING v11.0~\cite{szklarczyk2019string}  --- publicly available at \url{https://string-db.org/cgi/download.pl}  --- and BIOGRID v3.5.182~\cite{stark2006biogrid,oughtred2019biogrid}  --- publicly available at \url{https://downloads.thebiogrid.org/BioGRID/Release-Archive/BIOGRID-3.5.182/}). For a consistent analysis, all protein names and aliases have been standardized to follow the common nomenclature of official symbols of NCBI gene database (\url{ftp://ftp.ncbi.nlm.nih.gov/gene/DATA/GENE_INFO/Mammalia/} (Accessed: 28/03/2020)~\cite{murphy2019gene}). In the following we will refer to this comprehensive network, in standardized format, as BIOSTR.

The virus-host interactions for 45 viruses are collected from the STRING database  --- publicly available at \url{http://viruses.string-db.org/}. We consider interactions of any type as long as their confidence (score) is equal or larger than $0.25$. For Influenza C, Yellow fever, Human Hepatitis A, Dengue virus type 4 and Swinepox virus, due to the unavailability of data with score $0.25$, we consider the minimum available value $0.4$, instead. For each virus, we record the targeted human proteins and build a virus-host interactome by merging this information with BIOSTR. However, for the subsequent analyses, which are focused only on the human interactome, we discard virus-virus interactions.

It is worth noting that to build the COVID--19 virus-host interactions, a different procedure had to be used. In fact, since the SARS-CoV-2 is too novel we could not find its PPI in the STRING repository and we have considered, instead, the targets experimentally observed in Gordon et al~\cite{Gordon2020}, consisting of 332 human proteins. The remainder of the procedure used to build the virus-host PPI is the same as before. See Fig.~\ref{fig:virus_families} for summary information about each virus. 

Figure~\ref{fig:virus_contribution} shows a visualization of the human interactome where proteins targeted by viruses are highlighted. Remarkably, viruses tend to preferentially target similar regions related to specific functions of the human interactome. In fact, based on our dataset, TP53 (\emph{Tumor Protein p53}, NCBI Gene ID: 7157) is the most targeted node: it is responsible for inducing changes in metabolism, DNA repair, apoptosis and cell cycle arrest, and its mutations are associated with several human cancers. Other relevant targets (See Fig.~\ref{fig:virus_contribution}) include GK (\emph{Glycerol Kinase}, NCBI Gene ID: 2710), an important enzyme contributing to  regulate metabolism and glycerol uptake, and its mutations are associated with glycerol kinase deficiency; TBP (\emph{TATA-box Binding Protein}, NCBI Gene ID: 6908), which composes the transcription factor IID, which coordinates the activities of more than 70 polypeptides to initiate the transcription by RNA polymerase II; TLR4 (\emph{Toll Like Receptor 4}, NCBI Gene ID: 7099), relevant for recognizing pathogens and activating innate immunity; STAT2 (\emph{Signal Transducer and Activator of Transcription 2}, NCBI Gene ID: 6773), acting as a transcription activator within the cell nucleus: it is likely that it contributes to block interferon-alpha response by adenovirus; PTGS2 (\emph{Prostaglandin-endoperoxide Synthase 2}, NCBI Gene ID: 5743), a key enzyme involved in the process of prostaglandin biosynthesis; IFIH1 (\emph{Interferon Induced with Helicase C domain 1}, NCBI Gene ID: 64135), encoding MDA5, an intracellular sensor of viral RNA responsible for triggering the innate immune response: it is fundamental for activating the process of pro-inflammatory response that includes interferons, for this reason it is targeted by several virus families which are able to hinder the innate immune response by evading its specific interferon response.

\heading{Contributions} AG, OA and SB performed numerical experiments and data analysis. MDD conceived and designed the study. All authors wrote the manuscript. 

\heading{Competing financial interests} The authors declare no competing financial interests.

\heading{Acknowledgements} The authors thank Vera Pancaldi for useful discussions.

\bibliographystyle{naturemag}

\begin{small}
\bibliography{biblio}
\end{small}

\end{document}